\documentclass{article}
\usepackage{dcase2026_techrep,amsmath,graphicx,url,times,booktabs, tabularx}
\usepackage{multirow}
\usepackage{amssymb}
\usepackage{pifont}

\title{A Multi-Stage Separation-and-Classification Framework Guided by Complementary Acoustic-to-Semantic Clues}

\name{Younghoo Kwon$^{1}$,
      Junwoo Park$^{1}$,
      Han Yin$^{1}$,
      Jung-Woo Choi$^{1}\sthanks{Corresponding Author}$,
      }
\address{$^1$ School of Electrical Engineering, KAIST, Daejeon, Republic of Korea \\
        \{k0hoo, park.junwoo, hanyin, jwoo\}@kaist.ac.kr\\          
 }

\begin{document}

\ninept
\maketitle

\begin{sloppy}

\begin{abstract}
This report describes the system proposed for the DCASE 2026 Challenge Task 4: Spatial Semantic Segmentation of Sound Scenes (S5).
Specifically, we develop a multi-stage framework in which each stage couples a separation model with a classification model.
The first stage performs source separation and classification directly on the multi-channel mixture. Its outputs are then propagated to the following stage as two complementary clues that progressively refine each target estimate: 
(i) an enrollment clue, the separated waveform itself, serving as a low-level acoustic reference; 
and (ii) a class clue, the predicted label encoded as a one-hot vector. The third stage reuses the second-stage outputs under the same scheme, forming an iterative self-guided refinement process. 
In addition, we use a fine-grained frame-level audio embedding from an audio encoder pretrained on a large audio corpus as an additional clue to further improve the audio separation performance.
On the test set, the proposed system achieves a CAPI-SDRi of 15.51 dB, a mixture accuracy of 71.09\%, and a source accuracy of 78.62\%; with an improvement of 7.02 dB, 10.38\%p and 8.22\%p compared with the challenge baseline, respectively.
\end{abstract}

\begin{keywords}
Audio Source Separation, Audio Classification, Duration-Based Augmentation, Temporal-FiLM
\end{keywords}

\section{Introduction}
\label{sec:intro}

Spatial Semantic Segmentation of Sound Scenes (S5) aims to detect and separate individual sound events from a multi-channel mixture. Given a recording of several directional sources, interfering sounds, and diffuse background noise, an S5 system must both recognize target classes and recover an isolated waveform for each detected source. S5 systems can be applied in various applications, such as auditory scene analysis.

Following DCASE 2025 Task 4~\cite{yasuda2025}, the S5 task continues in DCASE 2026 Task 4~\cite{yasuda2026} with two modifications that bring the benchmark closer to real acoustic conditions. 
First, a single mixture may now contain multiple sources of the same class simultaneously. For example, several people may speak at once. Second, a mixture may contain zero target events, so that the system must reliably decide when no target sound is present while still being exposed to background noise and interference. Both changes substantially increase the difficulty of the task and require systems to be robust to source-count uncertainty and label ambiguity.

As a foundation for this task, we build upon a multi-stage self-guided framework initially developed for DeepASA~\cite{lee2026deepasa} and DCASE 2025 Task 4~\cite{kwonself}. These baseline frameworks progressively refine separation and classification using internally generated clues. Especially, the framework proposed in~\cite{kwonself} employs a Universal Sound Separation (USS) model, DeFT-Mamba-USS, to decompose multi-channel mixtures into distinct sources, followed by a Single-label Classification (SC) model, M2D-SC, to predict their labels. The separated waveforms and predicted classes are then fed into a Target Sound Extraction (TSE) model, DeFT-Mamba-TSE, as clues to form a self-guided, iterative refinement loop.

The initial baseline~\cite{kwonself} faced notable inefficiencies in its classification model (M2D-SC). Specifically, fine-tuning the classifier on a relatively small dataset degraded the diverse representational capacity of the pretrained model. Furthermore, converting the separated waveforms into magnitude-only mel-spectrograms caused the loss of phase and fine-grained frequency information. To overcome these limitations, a Dual-Path Classifier (DPC) was proposed~\cite{kwon2025soundseparation}. The DPC directly utilizes the intermediate object features produced from the USS model, alongside semantic features extracted from a frozen pretrained M2D model. This approach preserves fine-grained frequency information that is lost during Mel-spectrogram projection. Concurrently, an SEC module~\cite{kwon2025soundseparation} integrates semantic embeddings from pretrained models (e.g., M2D or MGA-CLAP) into DeFT-Mamba-TSE to enrich the class clues. These embeddings are projected to match the dimensions of the one-hot class embedding vector and are added element-wise before being injected into the DeFT-Mamba-TSE blocks via FiLM layers.

\begin{figure*}
    \centering
    \includegraphics[width=1\linewidth]{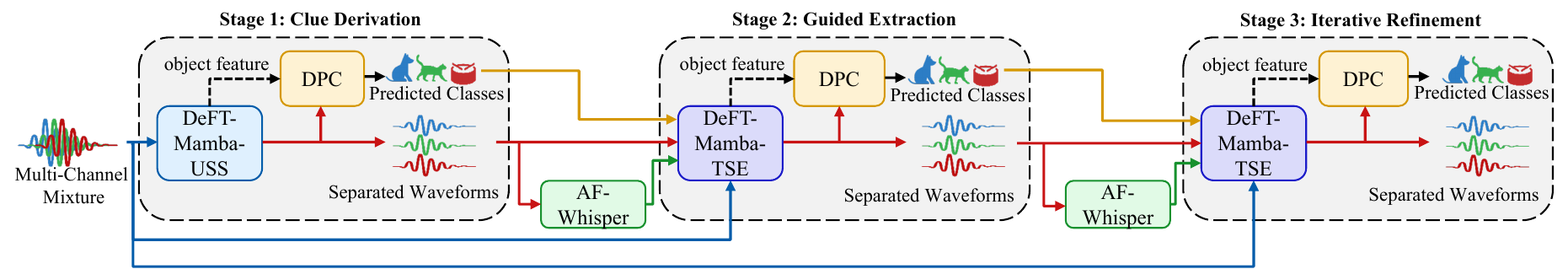}
    \caption{Overall architecture of the proposed multi-stage self-guided framework. Stage 1 derives initial clues using DeFT-Mamba-USS and M2D-DPC. Stages 2 and 3 iteratively refine target sound extraction using DeFT-Mamba-TSE, guided by the separated waveforms, one-hot predicted classes, and fine-grained AF-Whisper semantic embeddings.}
    \label{fig:overview}
\end{figure*}


Despite these advancements, applying the existing framework reveals two major limitations. First, the conventional SEC approach simply combines embeddings from one-hot class vectors and embeddings from a pretrained model through element-wise addition. This can dilute information contained in each embedding required for precise and rich target extraction. Furthermore, the embeddings from pretrained models have a coarse time resolution, failing to capture the fine-grained temporal dynamics. Second, we observe a classification performance degradation in percussive classes (e.g., percussion), which are frequently misclassified as silence. Our assumption is that this degradation is caused by the prevalence of extremely short, transient samples in the training dataset, which provide insufficient temporal context for reliable detection.
In this report, we propose two methods to overcome these limitations:
\begin{itemize}
    \item \textbf{Fine-grained Semantic Conditioning}: We utilize pretrained AF-Whisper~\cite{afwhisper} to obtain embeddings with a 20 ms time resolution to capture dense, frame-level temporal dynamics; and inject this embedding into the backbone through an independent Temporal Feature-wise Linear Modulation (Temporal-FiLM) layer~\cite{temporalFilm} to prevent information dilution.
    \item \textbf{Duration-based Augmentation} We introduce an augmentation strategy for difficult percussive classes by mixing short transient samples with longer sustained samples during training, thereby reducing the false-silence misclassification rate.
    \item \textbf{Class-specific Silence Threshold Optimization} To efficiently handle the silence prediction, we conduct a class-specific threshold tuning process to maximize the official metric, CAPI-SDRi.
\end{itemize}

In the official test set, our system achieves a CAPI-SDRi of 15.51 dB, a mixture accuracy of 71.09\%, and a source accuracy of 78.62\%, significantly outperforming the baseline by 7.02 dB, 10.37\%p, and 8.23\%p, respectively.

\section{Proposed Method}

\subsection{Framework Overview}
Figure~\ref{fig:overview} illustrates our proposed three-stage self-guided framework for joint separation and classification. In Stage 1 (Clue Derivation), DeFT-Mamba-USS decomposes the multi-channel mixture into distinct object features, which are directly processed by the DPC to predict class labels and detect silence. In Stage 2 (Guided Extraction) and Stage 3 (Iterative Refinement), DeFT-Mamba-TSE progressively refines the separated waveforms, followed by enhanced classification performance. Each refinement stage is guided by three complementary clues derived from the preceding stage: the separated waveform (enrollment clue), the predicted one-hot label (class clue), and a fine-grained semantic embedding extracted by AF-Whisper. The enrollment clues are concatenated with the multi-channel mixture along the channel dimension, and the class clues are injected through the FiLM \cite{Film} layers interleaved between the separation blocks.

\subsection{AudioFlamingo-whisper Embedding}
Although the features extracted by the pre-trained M2D model encode rich semantic information, they are derived from patchified representations with a temporal downsampling factor of 16, leading to relatively coarse temporal resolution and potentially limiting fine-grained temporal modeling.
Therefore, we employ a fine-grained frame-level embedding from the AudioFlamingo (AF)-Whisper encoder~\cite{afwhisper}, pretrained on a massive audio corpus. 

As shown in Figure~\ref{fig:injection}, this embedding is injected through additional Temporal-FiLM layers placed after each FiLM layer handling the one-hot vectors. This injection operates at a 20 ms time resolution to capture frame-level temporal dynamics. Let $c_{oh}$ denote the embedding from the one-hot class vector predicted by the DPC, and $E_{AF}$ represent the sequence of semantic audio embeddings extracted by AF-Whisper. To preserve definitive categorical information, $c_{oh}$ is injected into the intermediate feature maps of DeFT-Mamba-TSE via a standard FiLM layer, providing global static conditioning. Immediately following this, the frame-level embedding sequence $E_{AF}$ is injected through an additional Temporal-FiLM layer. This decoupled conditioning ensures that the TSE model benefits from dense temporal fluctuations for dynamic source tracking without compromising or overriding the strict categorical boundaries provided by the one-hot class clue.

\subsection{Class-specific Silence Threshold Optimization}
Since the acoustic characteristics vary between classes, the classification and silence detection performance also differ between classes. We observed that classes with highly percussive temporal characteristics (e.g., Percussion and Footsteps) or stationary acoustic patterns that resemble background noise (e.g., MechanicalFans) are more likely to be misclassified as silence. Furthermore, when the classifier predicts an incorrect class, the predicted silence score also tends to increase. To address this issue, we propose a class-wise silence threshold tuning strategy. 

For a given test set, we apply a class-specific silence threshold that maximizes CAPI-SDRi. We first initialize the silence threshold of 0.5. Then sweep the threshold for each class and select the value that yields the highest CAPI-SDRi. The tuned thresholds are applied only during inference and are not used during training. The proposed method improves CAPI-SDRi in two ways. First, it improves the detection accuracy of zero-target events, thereby reducing the relative impact of CAPI-SDRi penalties. Second, when the classifier produces an incorrect class prediction, the tuned threshold increases the likelihood of assigning the output to silence instead. As a result, error cases that would otherwise incur both false-negative and false-positive penalties are converted into cases containing only a false-negative, leading to a higher CAPI-SDRi.
\begin{figure}
    \centering
    \includegraphics[width=1\linewidth]{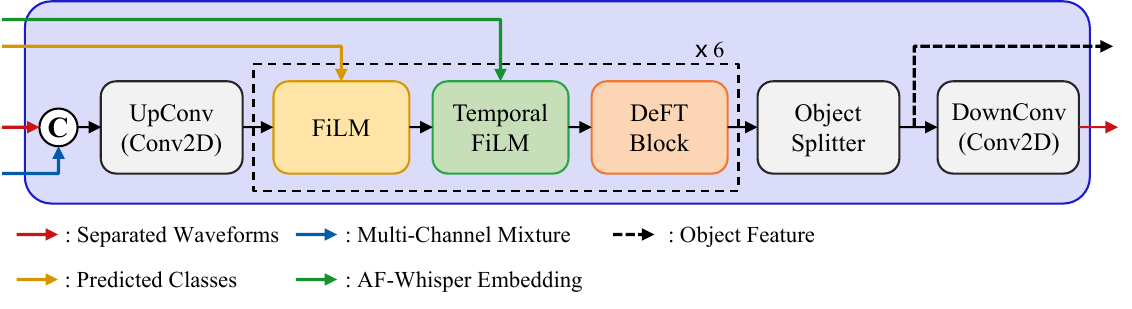}
    \caption{Architecture of the proposed DeFT-Mamba-TSE.}
    \label{fig:injection}
\end{figure}

\subsection{Duration-based Augmentation}

The instability in detecting percussive sounds primarily stems from training instances containing only a single, brief transient event (e.g., an isolated drum hit), which lacks sufficient temporal context for the classification module to reliably distinguish it. Therefore, to mitigate the frequent misclassification of percussive sounds as silence, we propose a duration-based augmentation technique applied during the dynamic synthesis of the training data.

Specifically, we establish a duration threshold $T_{th}$ for all samples belonging to percussive classes. During the batch generation process, instead of randomly sampling a single percussive audio file, we select two distinct files: a long sample $x_{long}$ with an active sound duration equal to $T_{th}$, and a short sample $x_{short}$ with a duration less than $T_{th}$. These two samples are stochastically mixed within the same spatial scene to form an augmented percussive source. By consistently exposing the network to percussive events with extended temporal presence, this augmentation drastically reduces the false-negative rate (silence misclassification) and enhances the extraction quality of short, impulsive acoustic events.

\section{Experimental Setups}

\subsection{Datasets and Augmentation}

To train our models, we dynamically generated 4-channel mixtures online for each sample using the SpatialScaper simulator~\cite{SpatialScaper}. 
For the duration-based augmentation, we set the threshold $T_{th}$ to 4 seconds.
While the foundational training data configuration follows the official DCASE 2026 Task 4 baseline~\cite{yasuda2026}, we introduced two modifications to improve data quality and inter-class discriminability:
\begin{itemize}
    \item Speech: To ensure the model learns from high-quality human speech representations, we replaced the provided speech dataset with the high-fidelity VCTK corpus~\cite{VCTK}.
    \item Vacuum Cleaner: Preliminary experiments revealed that the baseline dataset did not provide sufficient acoustic variance to effectively distinguish the VacuumCleaner class from the Blender class. To resolve this ambiguity, we added a newly curated subset from AudioSet-2M~\cite{audioset} to the baseline vacuum cleaner data~\cite{yasuda2026}. We strictly filtered this subset to include only high-quality samples annotated with a single "VacuumCleaner" label, minimizing label noise.
\end{itemize}

\subsection{Loss Functions}

Our multi-stage framework maintains a consistent loss function formulation across all training stages. At every stage, the network generates three distinct outputs for each estimated source: a separated waveform, a predicted class distribution (across the 18 target classes), and a scalar silence prediction. We optimize the network using a combined multi-task loss function defined as $\mathcal{L}_{total} = \mathcal{L}_{sep} + \mathcal{L}_{cls} + \mathcal{L}_{sil}$, where each component targets a specific output:
\begin{itemize}
    \item Separation Loss ($\mathcal{L}_{sep}$): The separated waveforms are optimized using the Source-Aggregated Signal-to-Distortion Ratio (SA-SDR) loss~\cite{SA-SDR} to ensure high-quality and phase-accurate source reconstruction.
    \item Classification Loss ($\mathcal{L}_{cls}$): The 18-class prediction is trained using the ArcFace loss~\cite{arcFace}. This explicitly maximizes inter-class separability and intra-class compactness.
    \item Silence Prediction Loss ($\mathcal{L}_{sil}$): The scalar output indicating the presence or absence of an active source is optimized using Binary Cross-Entropy (BCE) loss.
\end{itemize}



\subsection{Evaluation Metrics}
\label{ssec:evaluation_metrics}
To comprehensively evaluate both the separation quality and classification accuracy of our system, we employ three official metrics defined for DCASE 2026 Task 4: Class-Aware Permutation-Invariant Signal-to-Distortion Ratio improvement (CAPI-SDRi)~\cite{CApiSDRi}, Mixture Accuracy ($\mathrm{Acc}_{\mathrm{mix}}$), and Source Accuracy ($\mathrm{Acc}_{\mathrm{src}}$).

\textbf{CAPI-SDRi}: As the primary ranking metric for this task, it extends the standard CA-SDRi to jointly evaluate sound event detection and separation, explicitly accommodating conditions where multiple sources of the same class co-occur. It utilizes a permutation-invariant objective to resolve assignment ambiguity for same-class sources. Under this metric, incorrect predictions (false positives and false negatives) strictly receive a $0$ dB improvement penalty, while correct predictions contribute their permutation-optimized SDRi.

\textbf{Mixture-level Accuracy ($\mathrm{Acc}_{\mathrm{mix}}$)}: This metric evaluates classification performance on a per-mixture basis. A prediction is considered correct only if the entire set of predicted labels for a mixture exactly matches the set of ground-truth labels.

\textbf{Source-level Accuracy ($\mathrm{Acc}_{\mathrm{src}}$)}: This metric evaluates classification accuracy at the individual source level, measuring the proportion of correctly predicted labels among all separated foreground waveforms across the test set.

\section{Results and Discussions}

\subsection{Ablation Study}
To validate the effectiveness of our proposed methods across the multi-stage framework, we conducted a comprehensive ablation study on the development test set. As shown in Table 1, our final proposed system (Stage 3 equipped with both AF-Whisper and threshold tuning) achieves a CAPI-SDRi of 15.51 dB and a mixture accuracy of 71.09\%. This result significantly outperforms the official challenge baseline (CAPI-SDRi of 8.49 dB and $\mathrm{Acc}_{\mathrm{mix}}$ of 60.71\%), proving the superiority of our framework.

The performance variations in Table 1 highlight several key findings. First, the \textbf{iterative refinement} process is highly effective even without the proposed modules, as demonstrated by the steadily improved base network's CAPI-SDRi from Stage 1 (11.05 dB) to Stage 3 (14.43 dB). Second, the \textbf{metric-driven threshold tuning} plays a crucial role in handling zero-target conditions. For instance, in Stage 1, applying this tuning strategy improves the CAPI-SDRi from 11.05 dB to 11.64 dB. Finally, the \textbf{fine-grained semantic conditioning via AF-Whisper} yields a substantial leap in separation quality. In Stage 3, the introduction of the AF-Whisper embedding increases the Source Accuracy from 75.63\% to 76.09\%, which demonstrates that the dense temporal resolution successfully guides the DeFT-Mamba-TSE to accurately track dynamic sources. In addition, when synergized with threshold tuning, the performance is further improved and reaches a peak CAPI-SDRi of 15.51 dB. 

\begin{table}[t]
\small
\caption{Ablation study on the development test set. The hyphen denotes that the module is not applicable or not used in that stage. CAPI-SDRi is reported in dB, while Mixture Accuracy ($\mathrm{Acc}_{\mathrm{mix}}$) and Source Accuracy ($\mathrm{Acc}_{\mathrm{src}}$) are reported in \%.}
\label{tab:ablation}
\centering
\setlength{\tabcolsep}{4pt}
\resizebox{\columnwidth}{!}{%
\begin{tabular}{@{}cccccc@{}}
\toprule
\multirow{2}{*}{Stage} & \multirow{2}{*}{\begin{tabular}[c]{@{}c@{}}AF-\\Whisper\end{tabular}} & \multirow{2}{*}{\begin{tabular}[c]{@{}c@{}}Threshold\\Tuning\end{tabular}} & \multicolumn{3}{c}{Development Test Set} \\
\cmidrule(l){4-6}
 & & & CAPI-SDRi $\uparrow$ & $\mathrm{Acc}_{\mathrm{mix}}$ $\uparrow$ & $\mathrm{Acc}_{\mathrm{src}}$ $\uparrow$ \\
\midrule
baseline & - & - & 8.49 & 60.71 & 70.39 \\
\midrule
\multirow{2}{*}{1} & - & \ding{55} & 11.05 & 58.66 & 70.09 \\
  & - & \checkmark & 11.64 & 62.80 & 72.90 \\
\midrule
\multirow{4}{*}{2} & \ding{55} & \ding{55} & 13.43 & 64.09 & 72.26 \\
  & \ding{55} & \checkmark & 13.72 & 66.01 & 72.47 \\
  & \checkmark & \ding{55} & 14.03 & 64.02 & 73.55 \\
  & \checkmark & \checkmark & 14.26 & 65.48 & 74.44 \\
\midrule
\multirow{4}{*}{3} & \ding{55} & \ding{55} & 14.43 & 65.41 & 75.63 \\
  & \ding{55} & \checkmark & 15.36 & 71.16 & 78.64 \\
  & \checkmark & \ding{55} & 14.65 & 66.07 & 76.09 \\
  & \checkmark & \checkmark & \textbf{15.51} & \textbf{71.09} & \textbf{78.62} \\
\bottomrule
\end{tabular}%
}
\end{table}

\subsection{Class-specific Silence Threshold Optimization}

When the silence threshold is optimized per class for CAPI-SDRi, we observe two distinct trends, illustrated by the two representative classes in Figure~\ref{fig:threshold}. Note that the threshold on the $x$-axis represents the raw logit value. In the first trend (\textit{Pour}, left), performance is largely insensitive to the threshold near the default (t=0, black dashed) and degrades only at large thresholds, so the optimal threshold (red dashed) yields little gain over the default. Conversely, for the second trend (\textit{MechanicalFans}, right), performance is genuinely sensitive to the threshold and peaks at a substantially higher value, where the optimum clearly exceeds the default baseline.

For classes exhibiting this latter trend, a high threshold causes a larger fraction of outputs to be treated as silence, which is beneficial in two respects. First, when the model erroneously predicts an active source for a region that is in fact silent, a higher threshold correctly suppresses this output back to silence. Second, when the model assigns a region to an incorrect class, treating that output as silence removes a spurious detection (a false positive), avoiding the penalty it would otherwise incur in the CAPI-SDRi computation. Plotting curves analogous to Figure~\ref{fig:threshold} for all classes, we find that \textit{MechanicalFans}, \textit{Clapping}, \textit{Dishes}, \textit{Footsteps}, and \textit{Percussion} likewise require high thresholds.

\begin{figure}
    \centering
    \includegraphics[width=1\linewidth]{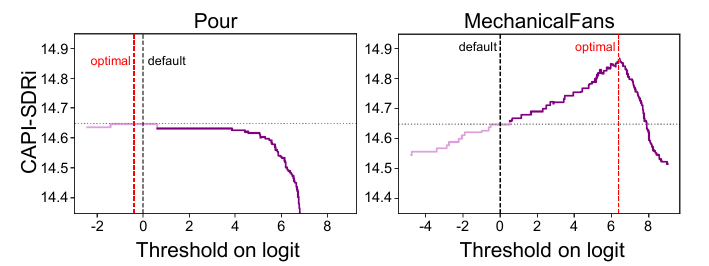}
    \caption{Impact of class-specific silence threshold tuning on CAPI-SDRi for the Pour (left) and Mechanical Fan (right) classes on the development test set.}
    \label{fig:threshold}
\end{figure}

\subsection{Effect of Duration-based Augmentation}
As observed during preliminary experiments, short and transient sound events are frequently missed by the classification module and incorrectly suppressed as silence. To explicitly evaluate the effectiveness of our duration-based augmentation in mitigating this issue, we analyzed the classification outcomes at the initial Clue Derivation stage (Stage 1) for three classes characterized by highly impulsive acoustics. Evaluating at Stage 1 is crucial because it serves as the foundational bottleneck of our multi-stage framework.

Table 2 compares the number of sample predictions before and after applying the augmentation strategy. For the \textit{Percussion} class, the network initially struggled with brief transients, misclassifying 74 samples as silence. After implementing the duration-based augmentation, this false-negative count drastically dropped to 28, while correct predictions surged from 47 to 76. Similarly, for the \textit{CupboardOpenClose} class, the augmentation resolved severe silence confusion, reducing `Silence' predictions from 64 to 13 and nearly doubling the `Correct' predictions (64 to 121). This detailed breakdown clearly demonstrates that our augmentation strategy reduces the false-silence rate for impulsive sounds.

\begin{table}[t]
\small
\caption{Effect of duration-based augmentation on the classification outcomes for transient sound classes in the development test set. The values represent the number of samples from the initial Clue Derivation stage (Stage 1) predicted as the correct class, a wrong class, or misclassified as silence.}
\label{tab:augmentation}
\centering
\setlength{\tabcolsep}{4pt}
\resizebox{\columnwidth}{!}{%
\begin{tabular}{@{}lcccccc@{}}
\toprule
\multirow{2}{*}{Class} & \multicolumn{3}{c}{Before Augmentation} & \multicolumn{3}{c}{After Augmentation} \\
\cmidrule(lr){2-4} \cmidrule(l){5-7}
 & Correct & Wrong & Silence & Correct & Wrong & Silence \\
\midrule
Percussion & 47 & 19 & 74 & 76 & 36 & 28 \\
Dishes & 55 & 17 & 64 & 68 & 48 & 20 \\
CupboardOpenClose & 86 & 11 & 64 & 121 & 27 & 13 \\
\bottomrule
\end{tabular}%
}
\end{table}

\section{Conclusions}

In this report, we presented a multi-stage separation-and-classification framework for DCASE 2026 Task 4. To tackle the increased complexity of the new benchmark, specifically source-count uncertainty and zero-target events, we advanced our previous architecture by integrating a Dual-Path Classifier (DPC) that directly utilizes object features. Furthermore, we introduced fine-grained semantic conditioning via AF-Whisper to provide dense temporal guidance without diluting categorical boundaries, and a duration-based augmentation strategy to prevent the critical loss of transient sounds. Coupled with a metric-driven threshold tuning during inference, our fully equipped system achieved a CAPI-SDRi of 15.51 dB and a mixture accuracy of 71.09\% on the development test set, demonstrating massive improvements over the challenge baseline. These results confirm that progressively refining internally generated multi-clues provides a highly robust and state-of-the-art solution for spatial semantic segmentation of sound scenes.

\section{ACKNOWLEDGEMENT}
This work was supported by the National Research Foundation of Korea (NRF) grant (No. RS-2024-00337945), STEAM research grant (No. RS-2024-00464269) funded by the Ministry of Science and ICT of Korea government (MSIT), and the BK21 FOUR program through the NRF grant funded by the Ministry of Education of Korea government (MOE).

\newpage

\bibliographystyle{IEEEtran}
\bibliography{refs}
%
%
%
%
%
%
%
%
%

\end{sloppy}
\end{document}